# Comparative Evaluation of Modern Deep Learning Methodologies for Portfolio Optimization

Samuel Ozechi[1] , Banjo A. Francis[2], Wisdom Yakanu[3], Joe Wayne Byers[4]

## Abstract

This study presents a broad portfolio optimization model that incorporates advanced deep learning architectures with traditional financial models to improve risk-adjusted performance. Historical data from 2015 to 2023 across a variety of asset classes were used, including equities, Exchange-Traded Funds (ETFs), and bonds. This research further examines the structural capabilities and predictive power of Graph Neural Networks (GNNs), Deep Reinforcement Learning (DRL), Transformers, and Autoencoders. The suggested models combine covariance estimation, forecasting return, dynamic asset allocation, and dimensionality reduction. Composite models such as Transformer+GNN and Autoencoder+DRL are also explored to leverage both relational and temporal structures in market data. Performance is evaluated through a robust backtesting method, comparing volatility, cumulative return, maximum drawdown, annualized return, and Sharpe ratio across seven strategies, among which are Equal-Weighted, 60/40 allocations, and Mean-Variance Optimization (MVO). Empirical outcomes show that composite learning models demonstrate better stability and risk control, with Transformer+GNN attaining the lowest volatility and drawdown. Mean-Variance Optimization with well-calibrated inputs recorded the maximum cumulative return and Sharpe ratio. Standalone DRL fell short of expectations due to limited structural awareness. The Autoencoder echoed the behaviour of Equal-Weight allocations, which underscores the significance of dynamic policy learning. The results obtained conform to existing literature highlighting the value of relational modeling and feature compression in financial applications. The result of Mean-Variance Optimization in this context challenges previous critiques by establishing that traditional methods retain competitiveness when paired with correct predictive inputs. This study supports the current discourse on data-driven portfolio construction and proposes that the incorporation of deep learning and financial theory can produce robust and adaptive investment strategies. Future studies may examine the application of latent financial representations within traditional optimization frameworks to further advance robustness and scalability.

---

[1]WorldQuant University (ozechisamuel@gmail.com)

[2]WorldQuant University (olufrank200@yahoo.com)

[3]WorldQuant University (ellinamwisdom@gmail.com)

[4]WorldQuant University (joe.byers@wqu.edu)

# 1. Introduction

Financial markets have experienced swift shifts as a result of technological disruption, globalization, and the growing multifaceted inter-asset dependencies. Traditional portfolio optimization methods, such as Mean-Variance Optimization, Capital Asset Pricing Model, and the Black-Litterman model, often depend on static covariance matrices and linear mean-variance assumptions making them unable to adapt in high-dimensional, non-stationary, and non-linear market settings. Traditional models rely largely on fixed assumptions and manually engineered features. Deep learning models are capable of automatically learning hierarchical representations from raw dataset, adapting to evolving market patterns as well as generalizing across market environments. Progressions in Deep learning architectures like Graph Neural Networks, Deep Reinforcement Learning, Transformers and Autoencoders have done better than legacy approaches in fields like bioinformatics, Natural Language Processing, and computer visions. The integration into quantitative finance for portfolio optimization problems remains limited, sparsely examined, fragmented, and underexplored. This study suggests a multi-model blended Artificial Intelligent model for portfolio optimization making use of Transformers for sequential temporal return modeling, Deep Reinforcement Learning (DRL) for dynamic weight learning, Graph Neural Networks (GNNs) for learning evolving inter-asset dependencies, Autoencoders for latent market factor discovery, Hybrid DRL+Autoencoder, and Transformer+GNN architectures for superior predictive power. We seek to answer: "Can blended deep learning models reliably surpass or exceed traditional portfolio strategies across a range of market regimes while remaining interpretable and applicable in large-scale institutional deployment?"

## 1.2 Objectives

The following research objectives would be addressed:
1. Build and train a Deep Reinforcement Learning (DRL) model to learn optimal portfolio allocations from historical data.
2. Develop an Autoencoder for dimensionality reduction and noise filtering of return streams used as input to DRL (DRL+Autoencoder hybrid).
3. Train a Graph Neural Network (GNN) to learn the evolving inter-asset correlation/covariance structure directly from data.
4. Implement a Transformer architecture to capture temporal dependencies in asset returns.
5. Integrate Transformer+GNN models as a hybrid model, where Transformer captures return sequences and GNN learns relationships among assets.
6. Benchmark against Equal-Weighted, 60/40, and MVO.
7. Evaluate risk-adjusted performance with Sharpe, Max Drawdown, and Turnover over historical backtests.

## 1.3 Research Questions

The following research questions were raised:
1. Does adding latent factors (Autoencoder) stabilize DRL training and improve out-of-sample Sharpe ratio?
2. Can a Transformer+GNN hybrid better predict covariance and return matrices than empirical estimation?
3. How does a DRL+Autoencoder hybrid compare to a pure DRL or Autoencoder model?
4. What is the interpretability tradeoff when using deep models vs. traditional optimizers?
5. Under what market regimes do AI models outperform benchmarks?

This work fills a crucial void in the literature by offering both empirical insight and methodological synthesis from a comparative perspective, pushing the frontier of intelligent portfolio design.

### 1.3 Background to the Study

The formalization of portfolio optimization began in 1952, when economist Harry Markowitz introduced Modern Portfolio Theory (MPT). His work offered a practical way to think about building investment portfolios by balancing risk and return. He argued that instead of just optimizing returns, investors should aim for the best possible trade-off by getting the most return for the level of risk they are willing to take. He established the idea of applying portfolio variance to measure the level of risk an investor can take, a theory that continues to remain instrumental in shaping how investors approach investing strategies even today. Mathematically, he defined optimization problem as:

Minimize:

$$w^T \Sigma w \tag{1}$$

Subject to: $\quad w^T \mu = p$
$\quad\quad\quad\quad\quad w^T 1 = 1$

where: $w$ = vector of portfolio weights,
$\Sigma$ = covariance matrix of asset returns,
$\mu$ = vector of expected returns,
$p$ = target portfolio return.

The findings lead to what is known as "the efficient frontier" which is an array of optimal portfolios given the maximum expected return for a specified level of risk. The real-world application of MPT depends on several important assumptions that simplify the difficulties of real-world market environments. MPT posits that asset returns are typically distributed and further assumes that investors are both rational and risk-averse, there are no transaction costs or taxes, and that MPT operates under the assumption of efficient markets. The theory assumes that the covariances and correlations amongst asset returns are static and stable over a period of time. Modern Portfolio Theory remains a foundation of modern finance and it has not been immune to critique. One of the most persistent disparagements centers on the supposition of normally distributed returns.

Empirical proof shows that financial returns frequently deviate from the normal distribution. It shows skewness and fat tails; a phenomenon known as excess kurtosis. Such deviations, as noted by Benoit Mandelbrot in 1963, propose an advanced likelihood of extreme market events than MPT accounts for. In unstable or crisis periods, asset correlations can change dramatically, compromising the reliability of historical estimates. Investigation by Ledoit and Wolf (2004) key points how this instability challenges the robustness of optimized portfolios based on historical datasets. Real-world asset managers often substitute MPT with robust optimization methods such as Bayesian models for more recent data-driven models that better account for non-stationary, time-varying, and market complexity. The Capital Asset Pricing Model (CAPM), proposed by William Sharpe (1964), John Lintner (1965), and Fischer Black (1972), signifies the critical and foundational moment in modern asset pricing theory. The concept was built on Markowitz's Modern Portfolio Theory by formulating a linear relationship between expected return and systematic risk as captured by the beta coefficient. CAPM postulates further that the expected returns on assets are determined by its sensitivity to the entire market.

The mathematical formula is as follows:
$$E(R_i) = R_f + \beta_i[E(R_m) - R_f] \tag{2}$$

where: $E(R_i)$ = expected return of the asset,
$R_f$ = risk-free rate,
$E(R_m)$ = expected return of the market portfolio,
$\beta_i$ = asset's beta, defined as $\beta_i = \frac{Cov(R_i, R_m)}{Var(R_m)}$

Equation (2) states that investors should only be rewarded for systematic risk and that idiosyncratic risk can be diversified away in a well-constructed portfolio allocation. Capital Asset Pricing Model (CAPM) splits risks into broad market risk which affects most investments and individual risks that are tied to specific assets. This aids in understanding how investments should be priced in relative to the overall market. The Security Market Line (SML) illustrates how the assets' expected return is connected to its sensitivity to market movements (beta) so that investors can judge whether an asset is priced fairly or objectively.

Black's Zero-Beta CAPM (1972) is one of the extensions and alterations of CAPM which relaxed the supposition of a risk-free asset. This results in a linear asset pricing model even in the absence of a risk-free rate and multi-factor models such as the Fama-French Three-Factor Model (1993) and Carhart Four-Factor Model (1997). This was subsequently expanded by the CAPM to address persistent irregularities by adding size, value, and momentum factors. CAPM is well-known for its theoretical clarity and influence; it has faced significant scrutiny collectively from both empirical research and practical application. Notably, is the work of Fama and French (1992) which established that the central model explanatory variable, beta, frequently fails to account for observed differences in asset returns. This has called into question the predictive model strength in real-world markets. Fischer Black and Robert Litterman (1992) proposed "The Black-Litterman Model" at Goldman Sachs to tackle some of the main challenges associated with traditional mean-variance optimization, mostly its instability, sensitivity to input parameters, and incoherence with investor views. They suggested a Bayesian method that starts with equilibrium returns implied by the Capital Asset Pricing Model which is known as implied returns or prior estimates, and further fine-tunes these returns using investor-specific views to give posterior (updated) expected returns. These updated returns are later used in mean-variance optimization which result in more stable, diversified, and portfolios that are interpretable.

$$\Pi = \delta * \Sigma * w_{mkt} \tag{3}$$

where: $\Pi$ = implied excess returns (from CAPM equilibrium),
$\delta$ = risk aversion coefficient,
$\Sigma$ = covariance matrix of returns,
$w_{mkt}$ = Market capitalization weights.

Under the assumption of views Q and a confidence matrix $\Omega$, the Black-Litterman model frequently updates the expected return vector using equation (4):

$$E(R) = [(\tau\Sigma)^{-1} + P^T\Omega^{-1}P]^{-1}[(\tau\Sigma)^{-1}\Pi + P^T\Omega^{-1}Q] \tag{4}$$

where: P = Pick matrix representing the assets involved in each view.
$\tau$ = Scalar representing the uncertainty in the prior,
Q = Vector of investor views on expected returns,
$\Omega$ = Diagonal matrix representing confidence in those views.

The Black-Litterman method suggests several main advantages that address significant limitations of traditional portfolio optimization techniques. One of these is its ability to integrate the views of the investors in a structured manner without entirely discarding the implied market equilibrium. This balance between subjective insights and objective market consensus gives rise to more realistic and stable expected returns. The model is very flexible and it allows investors to articulate both absolute and relative opinions on asset performance while stipulating their degree of confidence in those views. This flexibility makes the Black-Litterman model well-suited for real-world asset allocation in a multi-faceted, complex, and uncertain environment. The Black-Litterman model likewise subject to limitations:
1. The quality of output depends on how well the views and confidence levels of investors are stated.
2. Black-Litterman still depends on Gaussian assumptions akin to CAPM and MPT.
3. Adding many views to a large portfolio increases computational and modeling complexity.
4. It lacks flexibility to non-linear relationships, dynamic and time-varying behaviors, particularly in fast-moving or high-frequency markets.

The Black-Litterman model remains highly crucial in both academic literature and institutional or established portfolio management. In essence, the traditional portfolio optimization models (see Table 1) have contributed an essential role in financial theory and practice. These models have been extensively implemented in academia, asset management, and regulatory frameworks. Their real-world application shows a set of common structural and practical limitations:

| Model | Key Innovation | Main Limitation |
| --- | --- | --- |
| MPT | Mean-variance optimization | Instability from return estimates, static assumptions |
| CAPM | Systematic risk pricing via beta | Empirical failure of beta to explain returns |
| Black-Litterman | Bayesian blending of market views | Requires strong subjective inputs, lacks nonlinearity |

Table 1. Summary of the Traditional Portfolio

The rising in complexity of modern markets such as algorithmic strategies, high-frequency trading, and global interdependencies subject traditional models struggle to:
1. Capture non-linear dependencies among assets and macroeconomic variables.
2. Incorporate time-varying features such as changing volatilities, correlations, and risk regimes.
3. Integrate unstructured or high-dimensional data, including news sentiment, ESG metrics, and social media signals.
4. Adapt dynamically to new market conditions or rare events: COVID-19 shock, financial crises.

The challenges mentioned have led to a paradigm shift in portfolio research and practice, where data-driven and machine learning approaches are increasingly viewed as capable alternatives. The growing literature (e.g., Heaton et al., 2017; Gu, Kelly & Xiu, 2020; Jiang et al., 2017) demonstrates that machine learning models frequently outperform traditional benchmarks in both predictive accuracy and out-of-sample portfolio returns, particularly in complex, volatile environments.

## 1.4 Machine Learning in Portfolio Optimization

Machine learning has improved the way investors approach portfolio allocation, mostly because it doesn't have the same restrictions as traditional methods and can uncover useful relationships and patterns from huge sets of financial data. Nowadays, there's a broad range of machine learning techniques, including simple approaches that learn from labeled data to more complex ones like deep learning and reinforcement learning.

## 2 Theoretical Framework

Deep learning (DL) has revolutionized several domains by enabling models to learn hierarchical representations from raw data. Application of DL models in finance offer powerful tools to manage and handle high-dimensional, noisy datasets, non-linear relationships between market variables, and temporal dependencies in financial time series along with multi-modal input sources like price, news, sentiment, and fundamentals. Deep Learning methodologies can without manual intervention learn features, handle sequential data through recurrent structures, and even model inter-asset relationships leveraging graph-based methods. These abilities make them particularly attractive for tasks such as risk modeling, asset pricing, and dynamic portfolio allocation. This unit centers on four advanced deep learning methodologies such as Transformers, Autoencoders, Graph Neural Networks (GNNs), and Deep Reinforcement Learning (DRL).

### 2.1 Autoencoders

The Autoencoders (AEs) are a group of neural networks used mostly for unsupervised learning such as dimensionality reduction, denoising, and latent representation extraction. Figure 1. illustrates the basic structure of a standard autoencoder architecture. Many researchers have explored the implementation of autoencoders in financial markets. Fischer and Krauss (2018) included LSTM-based autoencoders for time-series compression before equity return forecasting, whereas Bao et al. (2017) applied Autoencoders for multi-step financial time series prediction using hybrid LSTM-Denoising Autoencoder (LSTM-DAE) structures. Autoencoder consists of two components, namely;

1. Encoder: is a function that maps the inputs $X \in R^d$ to a latent space representation $Z \in R^k$ where k < d.

2. Decoder: is a function that maps the latent codes **z** back to a reconstructed input $\widehat{X} \in R^d$

Mathematically, this can be expressed as:

1. **Encoding Function:**

$$Z = f_\theta(X) \qquad (5)$$

where $f_\theta$ is parameterized by weights θ.

2. **Decoder Function:**

$$\widehat{X} = g_\phi(Z) \qquad (6)$$

where $g_\phi$ is parameterized by weights ϕ.

**Objective Function**

The goal of training an autoencoder is to find parameters $\theta, and\ \phi$ that minimize the reconstruction loss between the input $X$ and the output $\widehat{X}$.

The reconstruction loss can be defined as:

$$L_{rec}(X, \widehat{X}) = ||X - \widehat{X}||_2^2 \qquad (7)$$

The training problem can be formulated as:

$$\min_{\theta,\phi} E_{X \sim p_{data}}[|| X - g_\phi(f_\theta(X)) ||_2^2], \qquad (8)$$

where: $p_{data}$ represents the true data distribution.

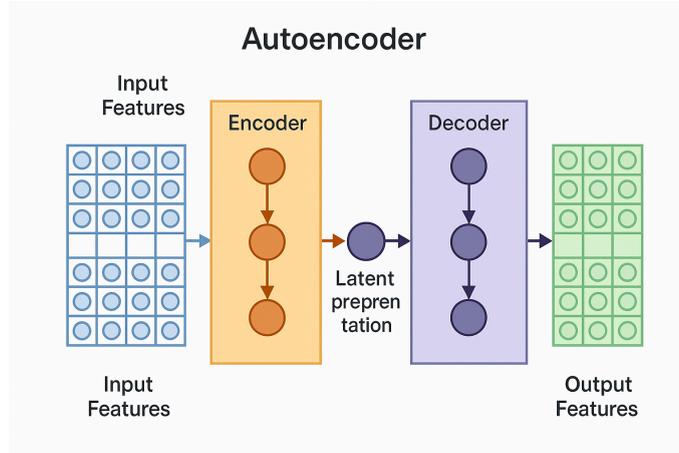

Figure 1. Autoencoder Architecture Diagram

Encoders and decoders are implemented as multi-layer perceptrons (MLPs) or as a convolutional neural network (for image data). Nonlinear activation functions such as ReLU and sigmoid are employed to introduce nonlinearity. Empirical studies have shown that portfolios constructed from AE-compressed features outperform those built on traditional statistical techniques with better risk-adjusted returns through effective feature extraction, improved diversification from latent factor discovery, and enhanced robustness to noise and market shifts.

**2.2  Graph Neural Networks**

Financial markets are inherently relational systems, where the performance, risks, and dependencies of assets are interlinked. As mentioned earlier, traditional models often struggle to explicitly capture these structural relationships between financial entities, whether it's co-movements in returns, sectoral affiliations, or macroeconomic exposures. Graph Neural Networks (GNNs) (see Figure 2.), offer a breakthrough in modeling such interdependencies. They treat financial systems as graphs, with nodes representing assets and edges encoding relationships such as correlation, causality, or supply-chain links. By learning over graph structures, GNNs can:
1. capture spatial dependencies in behavior or exposure.
2. generalize to unseen structures including new market settings or incorporating more assets.
3. likewise preserve relational inductive biases.

For every node in a Graph Neural Networks, the model collects features from its neighbors' nodes, updating its own representation through message passing. This appeared as a powerful tool in financial modeling by allowing researchers to include inter-asset relationships into portfolio optimization. The key contributors in this field include Wang et al. (2021) who applied Graph Convolutional Network (GCNs) for stock return prediction and the results demonstrated superior performance over conventional time-series models. Jiang et al. (2021) also applied Temporal Graph Attention Networks (TGANs) to dynamically evolving financial graphs for asset selection. Furthermore, Liu et al. (2022) extended GNNs to knowledge graphs in order to measure firm-level risks and inform portfolio decisions. Their studies emphasized the ability of Graph Neural Networks to capture spatial and temporal market structure, incorporate various data modalities, support more nuanced, interconnected decision-making, and offer a significant advancement over models that treat assets as independent entities. A graph is defined as:

$$\mathcal{G} = (\mathcal{V}, \mathcal{E}) \tag{9}$$

where: $V$ = set of nodes (vertices),
$\mathcal{E} \subseteq \mathcal{V} \times \mathcal{V}$ = set of edges (connections between nodes).
Each node $v \in \mathcal{V}$ can have an associated feature vector $X_v \in R^d$.

Message passing is the main idea of GNNs where every node updates its representation by collecting information from its neighbors. At every layer $l$, each node $v$ updates its hidden state $h_v^{(l)}$ based on its own prior state as well as the states of its neighbouring nodes $\mathcal{N}(v)$. The training of GNNs minimizes the supervised loss like cross-entropy in classification tasks. The equation of supervised loss is given as follows:

$$\mathcal{L} = \frac{1}{|\mathcal{V}_l|} \sum_{v \in \mathcal{V}_l} CrossEntropy(y_v, \hat{y}_v) \tag{10}$$

where: $\mathcal{V}_l$ = set of labeled nodes,
$y_v$ = true label,
$\hat{y}_v$ = predicted label.

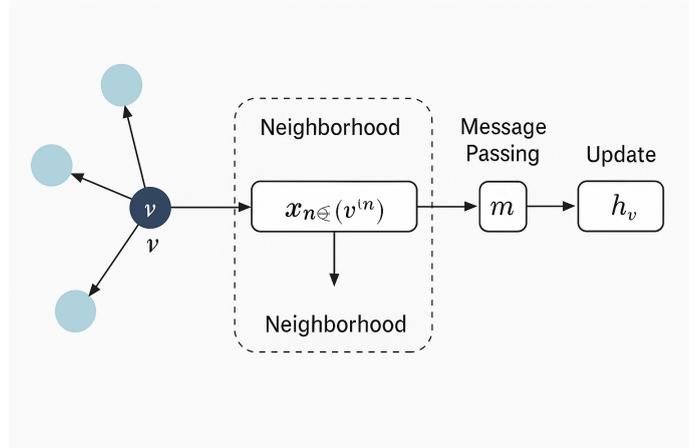

Figure 2: GNN Architecture with nodes, neighbourhoods, message passing, and update

## 2.3 Deep Reinforcement Learning (DRL)

Deep Reinforcement Learning (DRL) model combines reinforcement learning (RL) with deep learning (DL) algorithm (see Figure 3.) where agents learn to make decisions through interaction with their environments. Deep learning enables the agent to learn complex policies from high-dimensional inputs and an agent in RL learns a policy (a|s) that maps s states to actions to maximize a cumulative reward R. This reward can be modified to represent risk-adjusted performance, returns, Sharpe ratio, or other investment objectives. DRL creates this paradigm expandable or adaptable to the complex, continuous, and noisy environments of financial markets. The agent constantly observes market states such as price movements, macro variables, volatility and learns an optimal re-allocation strategy through experimental and error. Liang et al. (2018) employed Deep Reinforcement Learning to manage portfolios through a Deep Deterministic Policy Gradient (DDPG) algorithm, the result outperformed traditional benchmarks like the mean-variance approach. Zhang et al. (2020) also established how actor-critic algorithms could adapt to diverse asset universes and macroeconomic cycles, thereby capturing momentum and mean-reversion behaviors in a unified policy.

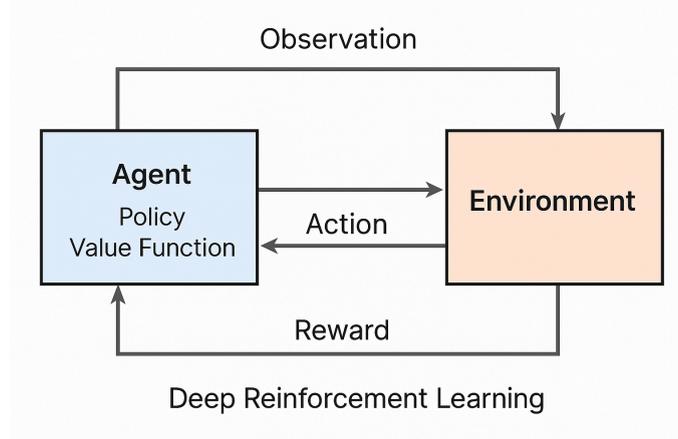

Figure 3. Deep Reinforcement Learning Architecture

Deep reinforcement learning architectures, models decision-making as a Markov Decision Process (MDP) and is defined as a tuple:

$$(S, A, P, r, \gamma) \tag{12}$$

where: $S$ = the set of states,
$A$ = the set of actions,
$P(s'|s, a)$ = the state transition probability,
$r(s, a)$ = the reward function,
$\gamma \in [0, 1]$ = the discount factor.

The purpose of this is to discover a policy $\pi(a|s)$ that may optimize expected cumulative reward over a period of time. The two key functions in reinforcement learning are the state-value function and the action-value function. The State-Value Function can be written mathematically as:

$$V^\pi(s) = \mathbb{E}_\pi[\sum_{t=0}^{\infty} \gamma^t r(s_t, a_t) | s_0 = s] \tag{13}$$

This denotes the expected return starting from state $s$ following policy $\pi$.
Mathematically, the Action-Value Function (Q-function) is written as;

$$Q^\pi(s, a) = \mathbb{E}_\pi[\sum_{t=0}^{\infty} \gamma^t r(s_t, a_t) | s_0 = s, a_0 = a] \tag{14}$$

Instead of explicitly storing tabular representations for value function **V** and the action-value function **Q** in Deep Learning Reinforcement (DRL), these functions are approximated using deep neural networks. This process is called value function approximation:

$$V_\theta(s) \approx V^*(s) \tag{15}$$

while Q-function approximation is given by;

$$Q_\theta(s, a) \approx Q^*(s, a) \tag{16}$$

The $\theta$s are the parameters of the deep network and the policy networks directly model the policy as policy approximation. This is expressed as follows:

$$\pi_\theta(a|s) \approx \pi^*(a|s) \tag{17}$$

The theory behind deep reinforcement learning rests solidly on the Bellman equations, value iteration, policy gradients, and theory of function approximation. Application of Deep Learning

Reinforcement (DRL) to portfolio optimization has introduced a set of complex capabilities that extend beyond traditional approaches. The ability of the model to perform dynamic portfolio rebalancing in response to evolving market regimes, thereby facilitating adaptive asset allocation based on shifting macroeconomic and financial conditions. DRL models can be explicitly designed to include transaction costs, slippage, trading fees, and execution latency directly within the learning paradigm. This leads to the development of strategies that are not only theoretically sound but also practically implementable in real-world trading environments. DRL supports multi-objective optimization and allows practitioners to balance competing goals such as maximizing returns, minimizing portfolio risk, and adhering to environmental, social, and governance (ESG) constraints. Beyond asset allocation, DRL has also proven effective in market-making and execution strategies where it models liquidity conditions and order flow dynamics as integral components of the state space. Deep Reinforcement Learning (DRL) offers considerable potential in financial applications; however, it presents several inherent challenges. It demands large volumes of data and significant computational resources which limit its applicability in data-constrained environments. DRL algorithms are often sample-inefficient, necessitating extensive synthetic data to reach policy convergence. The exploration-exploitation trade-off further introduces additional layers of complexity and deployment risks, as agents may engage in suboptimal decisions while seeking to improve policies.

## 2.4 Transformers for Portfolio Optimization

The Transformer architecture (Figure 4.), first proposed by Vaswani et al. (2017) in "Attention is All You Need", marked a paradigm shift in sequence modeling by replacing recurrence structure with self-attention mechanisms. This fundamental shift enables parallel computation and more effective capture of long-range temporal dependencies. Transformer was originally developed for natural language processing (NLP) tasks, it has increasingly been adopted in the realm of time series analysis, particularly in finance application where understanding temporal and contextual dependencies is essential. Transformer models are remarkable by their ability to handle complex sequential structures and large volume data across both time and sequences. Their ability to model both short and long-range across multiple sequences enhances their usefulness in asset return forecasting, risk modeling, and portfolio allocation. Recent study has successfully adapted Transformer-based models to financial forecasting, risk modeling, and asset allocation. Zhang et al. (2022) introduced the Financial Transformer (FinFormer) to effectively model both inter-asset correlations and temporal dependencies. This achieved a state-of-the-art performance in asset return prediction across US equities. Wu et al. (2021) proposed the Temporal Fusion Transformer (TFT), a model designed for interpretable and multi-horizon financial forecasting, valuable for rolling portfolio rebalancing decisions. Lim et al. (2021) demonstrated that attention-based models could outperform traditional LSTMs in predicting price movements across multiple assets simultaneously. In a transformer, each input token (word, asset price, time series vector, etc.) is embedded into a dense vector and combined with positional encodings to retain order information.

Let:

$$x = (x_1, x_2, ..., x_T) \in R^{T \times d_{model}}: \text{sequence of input embeddings} \qquad (18)$$

$$PE(i) \in R^{d_{model}}: \text{positional encoding at position } i.$$

The input to the encoder is:

$$z_i^0 = x_i + PE(i), \quad for \quad i = 1,...,T \qquad (19)$$

The core of the Transformer is the scaled dot-product attention, which computes a weighted sum of values, where the weights are derived from the similarity between queries and keys. Given as:

$$Queries\ Q \in R^{T \times d_k}$$
$$Keys\ K \in R^{T \times d_k}$$
$$Values\ V \in R^{T \times d_v}$$

The attention mechanism is defined as:

$$Attention(Q, K, V) = softmax\left(\frac{QK^T}{\sqrt{d_k}}\right)V \qquad (20)$$

Instead of performing a single attention function, the Transformer employs multi-head attention to allow the model to jointly attend to information from different representation subspaces:

$$MultiHead(Q, K, V) = Concat(head_1, ..., head_h)W^O \qquad (21)$$

where each head is computed as:

$$head_i = Attention(QW_i^Q, KW_i^K, VW_i^V) \qquad (22)$$

with learnable projections:

$$W_i^Q \in R^{d_{model} \times d_k}$$
$$W_i^K \in R^{d_{model} \times d_k}$$
$$W_i^V \in R^{d_{model} \times d_v}$$
$$W^O \in R^{hd_v \times d_{model}}$$

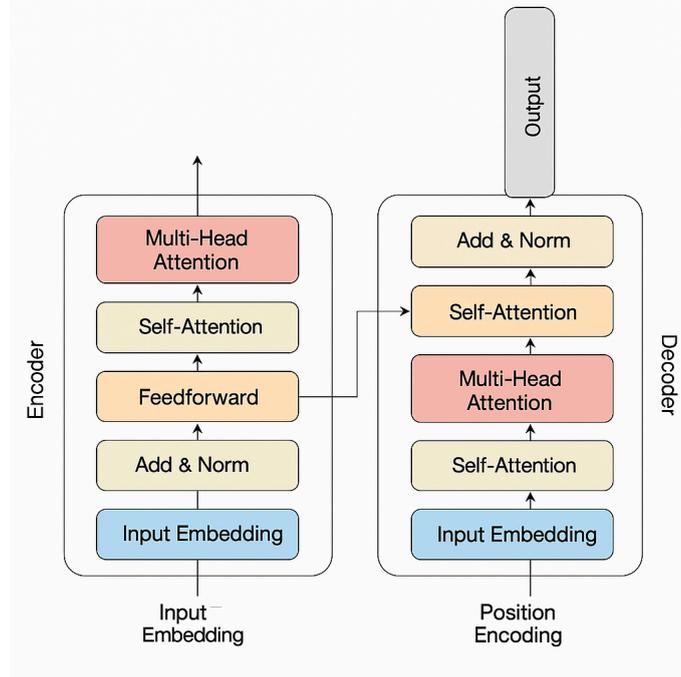

Figure 4. Transformer Architecture with encoder, decoder, multi-head attention, and feedforward blocks

Transformer architectures have emerged as a promising tool in various stages of the portfolio optimization process. This is due to their capacity to model complex temporal and cross-sectional dependencies. One application area is return forecasting, where transformers are employed to predict future asset returns. It achieves this by capturing patterns in historical price and volume data, macroeconomic variables, or alternative data sources. A further application lies in integrated asset selection and capital allocation. The attention mechanism enables the model to identify and emphasize assets with high return potential, thereby informing more efficient portfolio construction decisions. Transformers models typically require large volumes of data to achieve robust generalization and mitigate risk of overfitting. This is a critical concern in finance where features are often sparse, noisy, and prone to instability. Recent advancements to improve interpretability such as the use of attention heatmaps, the decision-making processes of Transformers largely opaque, posing ongoing challenges for explainability, model transparency, and regulatory compliance. Transformers have seen remarkable success in fields like natural language processing and computer vision, their application within financial contexts remains relatively emerging. This underscores the pressing need for domain-specific evaluation models, benchmark datasets, and the establishment of best practices tailored to financial modeling. The ability of transformers to capture complex temporal dependencies, integrate cross-sectional attention, and scale efficiently positions them highly appropriate for real-world portfolio management. However, their broader adoption in finance is still emerging and further empirical research is needed to benchmark their performance against other models in terms of robustness, generalizability, and financial interpretability.

## 3. Methodology

This project aims to evaluate portfolio optimization models that integrate advanced deep learning architectures such as Deep Reinforcement Learning, Autoencoder, Graph Neural Networks, Transformer and their hybrid models with benchmark optimization strategies such as Markowitz MVO. The purpose is to construct efficient portfolios that minimize risk while at the same time maximize returns. The research also combines dynamic rebalancing, risk management constraints, and transaction costs considerations to ensure the robustness of the optimization process and simulate real-world trading conditions.

### 3.1 Data Collection, Preprocessing, and Feature Engineering

This project uses historical daily data prices obtained through Yahoo Finance Application Programming Interface (API) from 2015 to 2023. The dataset includes a variety of assets which include stocks, ETFs, and bonds. The assets selected encompasses AAPL (Apple), TLT (US Treasury Bonds), AMZN (Amazon), SPY (S&P 500 ETF), EFA (Developed Markets ETF), GLD (Gold), and MSFT (Microsoft). Daily returns for each asset are calculated as the percentage change in asset prices over time. The dataset is enhanced with additional technical indicators such as measure of volatility, moving averages, and momentum. All features are also standardized to ensure that all variables are on a comparable scale which helps models' stability and improve the performance of machine learning algorithms during training.

### 3.2 Model Development

A variety of deep learning models are designed to evaluate their effectiveness to optimize returns and reduce risk which are the main reasons for portfolio optimization. The models include:
- Deep Reinforcement Learning (DRL) for portfolio allocation,
- Graph Neural Network (GNN) for covariance estimation,
- Transformer for return prediction,
- Autoencoder for dimensionality reduction and covariance estimation,
- and hybrid models' architecture combine to leverage their complementary strengths.

### 3.2.1 Deep Reinforcement Learning for Portfolio Allocation

The Deep Reinforcement Learning model is trained to learn portfolio weights optimally by interacting with historical financial dataset (asset returns) in order to maximize the Sharpe ratio. The DRL Trainer class is employed to train the model with hyperparameters through a systematic tuning process. Models' performance is evaluated based on multiple metrics like volatility, cumulative return, and Sharpe ratio to ensure stability and profitability in portfolio allocation.

### 3.2.2 Transformer for Return Prediction

The Transformer model is employed to identify and capture temporal relationships together with long-range dependencies in asset returns and facilitate the prediction of future returns. The model is then trained to optimize predictive accuracy on unscaled asset returns. Performance is evaluated on the basis of benchmark performance metrics including the Sharpe ratio and cumulative return.

### 3.2.3 Graph Neural Network (GNN) For Covariance Estimation

The Graph Neural Network (GNN) model is used to estimate the covariance matrix between assets from raw returns data. It identifies and picks up relationships as well as dependencies between assets. The model is then trained on raw returns and its effectiveness is assessed on its performance within the Markowitz optimization framework.

### 3.2.4 Autoencoder for Dimensionality Reduction and Covariance Estimation

The autoencoder is responsible to ensure dimensionality reduction by encoding assets returns into a lower-dimensional latent space. From these latent features, estimates for the covariance matrix are generated. The model afterward is trained to minimize reconstruction error and its performance is evaluated based on its contribution to portfolio allocation outcomes.

### 3.2.5 Hybrid Models

Two hybrid modeling approaches are explored to evaluate their efficiency in portfolio allocation and risk mitigation management; Deep Reinforcement Learning + Autoencoder, and Transformer + Graph Neural Networks (GNN). The DRL + Autoencoder model incorporates latent features extracted from the autoencoder with a Deep Reinforcement Learning (DRL) architecture, taking advantage of dimensionality reduction and reinforcement learning for optimization of portfolio allocation. The Transformer + Graph Neural Networks (GNN) hybrid model integrates the return of the Transformer prediction abilities with the Graph Neural Networks' (GNN) estimation of covariance, thereby provide a comprehensive model to portfolio allocation optimization by accurately forecasting returns and estimating covariances. The two hybrid models are evaluated using standard performance metrics such as cumulative return and Sharpe ratio.

### 3.2.6 Portfolio Optimization

Portfolio construction is implemented using the Markowitz Mean-Variance Optimization (MVO) model, which aims to maximize the Sharpe ratio by adjusting portfolio weights based on predicted returns and estimated covariances. Forecasted returns are generated by Transformer and the covariance matrix generated by the Graphical Neural Network and Autoencoder to perform mean-variance optimization, thereby maximizing the Sharpe ratio by optimizing and rebalancing portfolio weights.

**Key Considerations:**

Several practical considerations are incorporated into the optimization method. Transaction costs are accounted for by penalizing high portfolio turnover to encourage cost-efficient allocation strategies. Risk constraints such as maximum volatility and maximum drawdown are included to ensure portfolios remain within predefined risk limits.

**Dynamic Rebalancing:**
The portfolio undergoes periodic rebalance based on the latest forecasts from the underlying models. This replicates real-world portfolio management practices where adjustments happen in response to market conditions and changes in asset predictions. Such a dynamic and adaptive method enables the portfolio to adapt with dynamic market environments and maintain optimal performance over time.

### 3.2.7   Backtesting
The Backtester class is implemented to simulate and evaluate the performance of various strategies using historical price data. This process enables the assessment of model efficiency within a controlled and simulated trading environment. The strategies backtested include:
- Deep Reinforcement Learning (DRL)
- Autoencoder + DRL
- Transformer + Graph Neural Network (GNN)
- Autoencoder
- Markowitz Mean-Variance Optimization (MVO)
- Equal-Weighted Portfolio
- Traditional 60/40 Allocation

**Backtesting Dynamics**
In the course of backtesting, performance of various portfolio optimization strategies is evaluated taking transaction costs, dynamic rebalancing, and risk constraints into consideration. The backtester repeatedly applies the portfolio weights generated by each model over time and evaluates portfolio returns. Key performance metrics include volatility, Sharpe ratio, cumulative return, and maximum drawdown.

**Strategy Evaluation**
The backtesting facilitates a comprehensive comparison between different portfolio allocation strategies. Hybrid models such as Transformer + GNN and Autoencoder + DRL are benchmarked against traditional approaches namely Markowitz MVO, Equal-Weighted, and 60/40 Allocation. By assessing these approaches under consistent historical conditions, the backtester gives constructive insights into the strengths, weaknesses, and practical effectiveness of each strategy. This ensures the selection of the most efficient strategy for optimization of the portfolio.

### 3.2.8   Evaluation Metrics
The performance of the portfolio strategies was measured through Annualized Return, Cumulative Return, Annualized Volatility, Sharpe Ratio, and Maximum Drawdown (MDD)
Let assume we have daily portfolio returns $R_t$ for days t = 1,2, ..., T.

**Cumulative Return:** Is used to measure the total percentage gain or loss on an investment over a specific period of time or throughout the entire backtesting period.
It is calculated using the following formula:

$$CR = \prod_{t=1}^{T}(1 + R_t) - 1 \qquad (23)$$

where:   $R_t$ = Portfolio return on day t.

T = Total number of days in the backtesting period.

**Annualized Return:** this is the average yearly return earned by the portfolio over time. The formula is expressed as:
$$AR = (1 + CR)^{\frac{N}{Number\ of\ Days}} - 1 \qquad (24)$$
where:  N = the number of trading days in a year and is 252 days for stocks,
Number of Days = the total number of days in the backtesting period.

**Annualized Volatility:** is a measure of the amount of risk in the return of assets over a year and it is expressed as a standard deviation. The equation:
$$AV = standardDeviation(R_t) \times \sqrt{N} \qquad (25)$$
where:  $standardDeviation(R_t)$ = standard deviation of the daily portfolio returns,
N = number (252) of trading days in a year.

**Sharpe Ratio:** This is used to evaluate the performance of an investment in contrast to a risk-free asset after adjusting for its risk. The equation is formulated as;
$$\text{Sharpe Ratio} = \frac{AR - R_f}{AV} \qquad (26)$$
where:  $AR$ = annualized portfolio return,
$R_f$ = annualized risk-free rate,
$AV$ = annualized portfolio volatility.

**Maximum Drawdown (MDD):** Is the largest peak-to-trough decline in the value of an investment before a new peak is attained. It measures the downside risk over a specific period. The equation:
$$MDD = max_{t1<t2}\left(\frac{PortfolioValue_{t1} - PortfolioValue_{t2}}{PortfolioValue_{t1}}\right) \qquad (27)$$
where:  $PortfolioValue_t$ = Value of portfolio at time $t$
The maximum is taken over all pairs of time points $t_1$ and $t_2$ such that $t_1$ comes before $t_2$.

## 4.    Results and Performance Evaluation

This section presents a detailed comparative analysis (see Figure 5 - Figure 11) of the backtested performance of seven (7) portfolio allocation strategies over an out-of-sample historical period. The methods evaluated include DRL (PPO), Autoencoder, Autoencoder+DRL, Transformer+GNN, and in addition to benchmark strategies which include Mean-Variance Optimization (MVO), Equal weight, and a 60/40 portfolio. Each strategy's efficiency is quantified using the following financial metrics: Cumulative Return (CR), Annualized Return (AR), Annualized Volatility (AV), Sharpe Ratio (SR), and Drawdown (MDD).

## 4.1 Portfolio Weights Allocation Visualizations

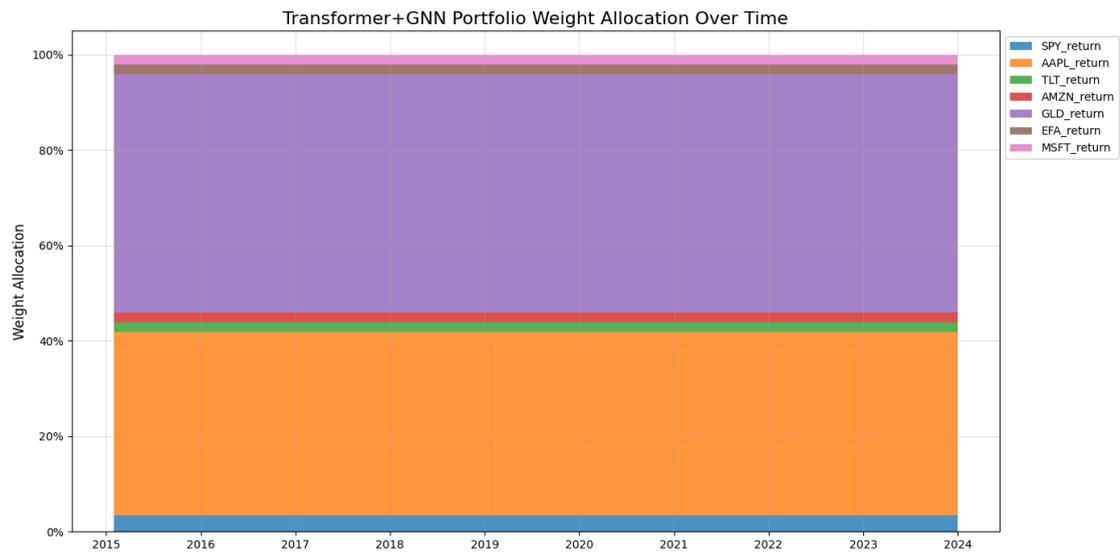

Figure 5. Transformer+GNN Portfolio Weight Allocation Over Time

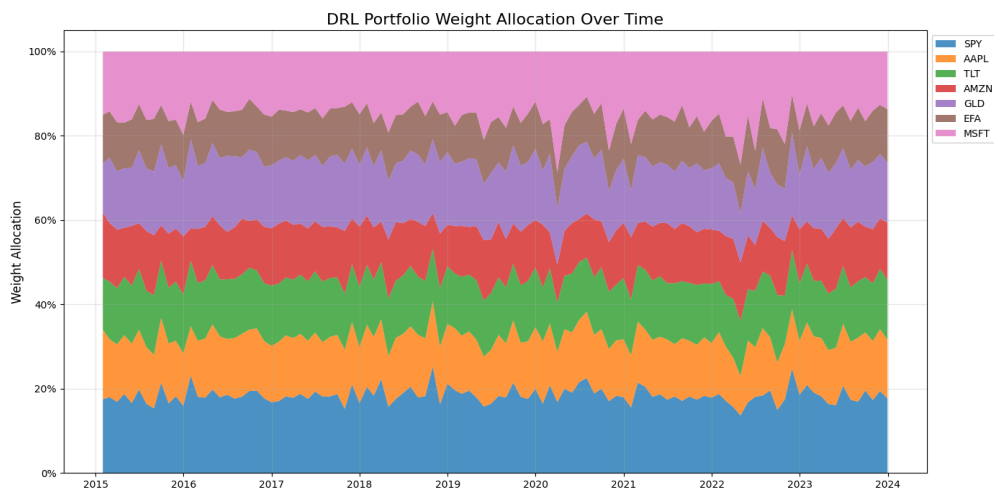

Figure 6. DRL Portfolio Weight Allocation Over Time

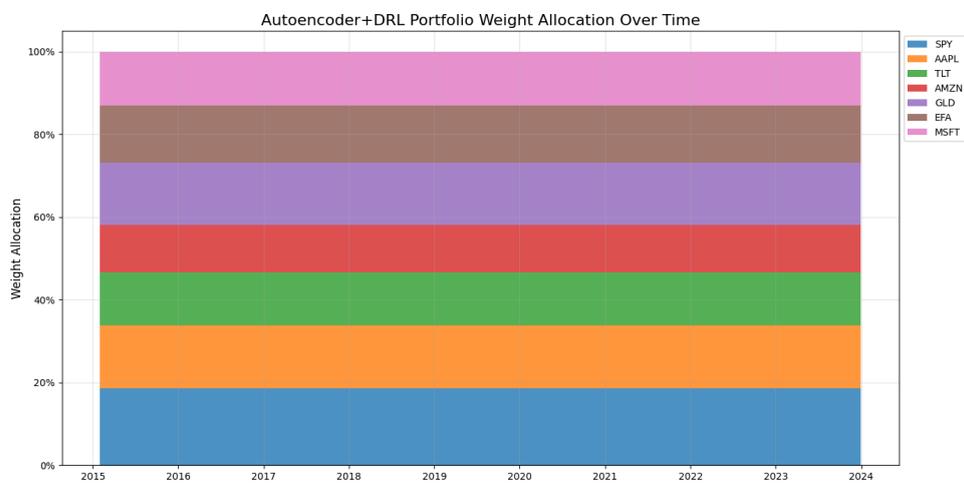

Figure 7. Autoencoder+DRL Weight Allocation Over Time

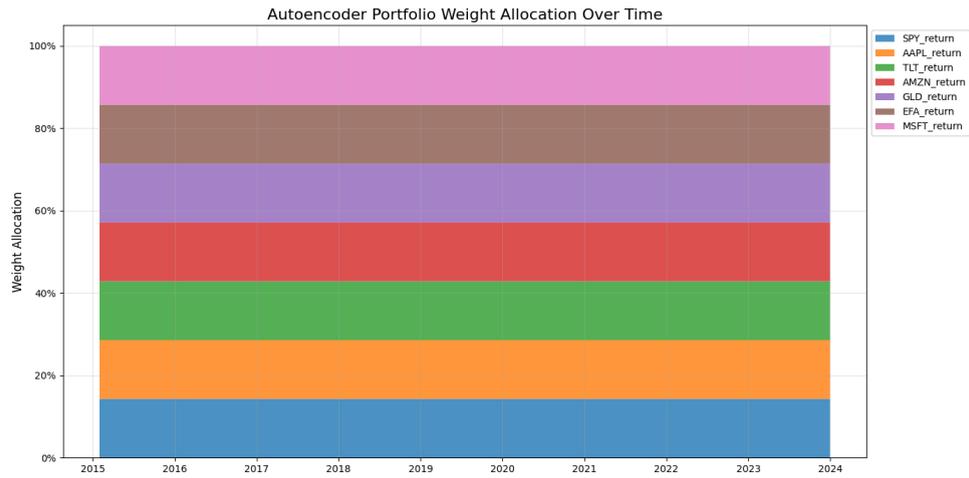
Figure 8. Autoencoder Portfolio Weight Allocation Over Time

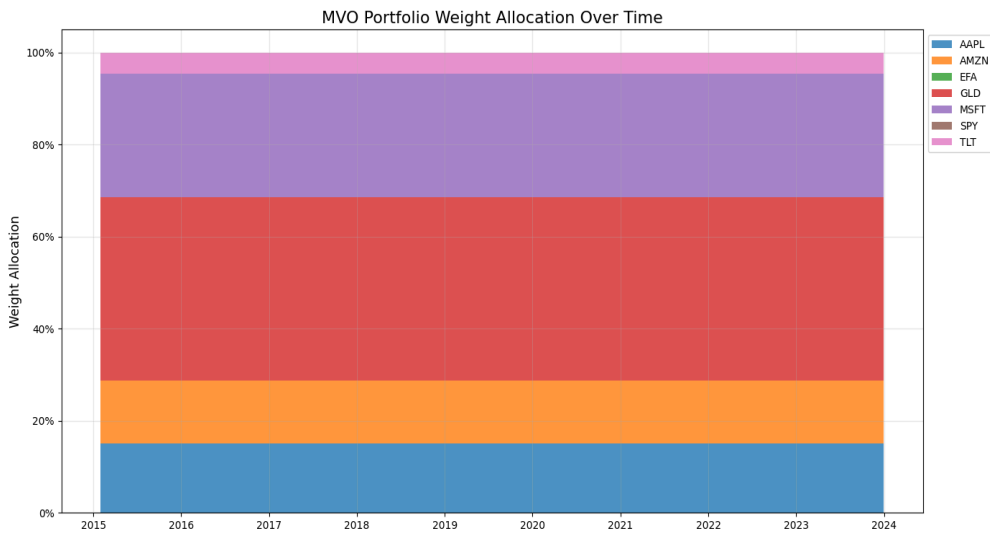
Figure 9. MVO Portfolio Weight Allocation Over Time

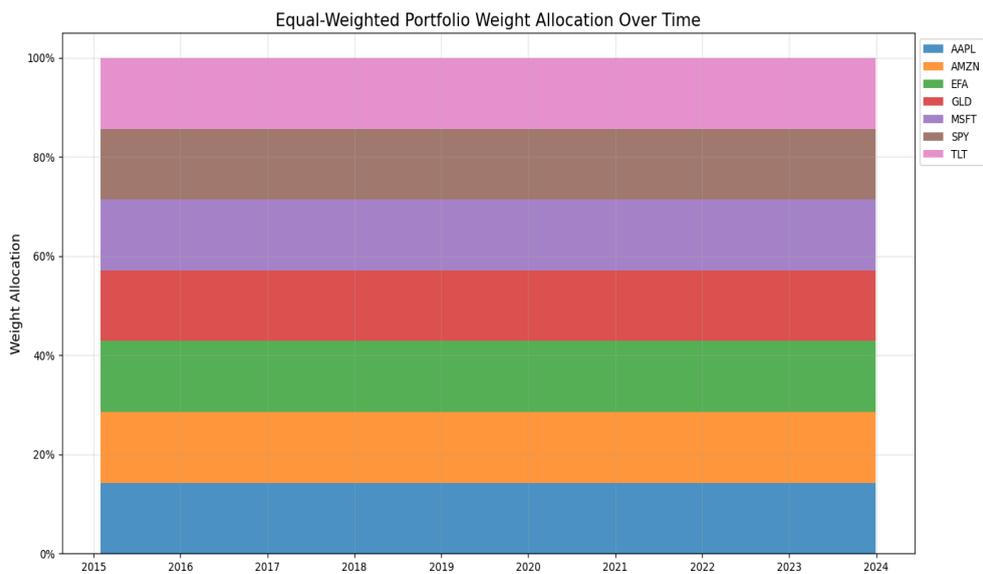
Figure 10. Equal-Weight Portfolio Weight Allocation Over Time

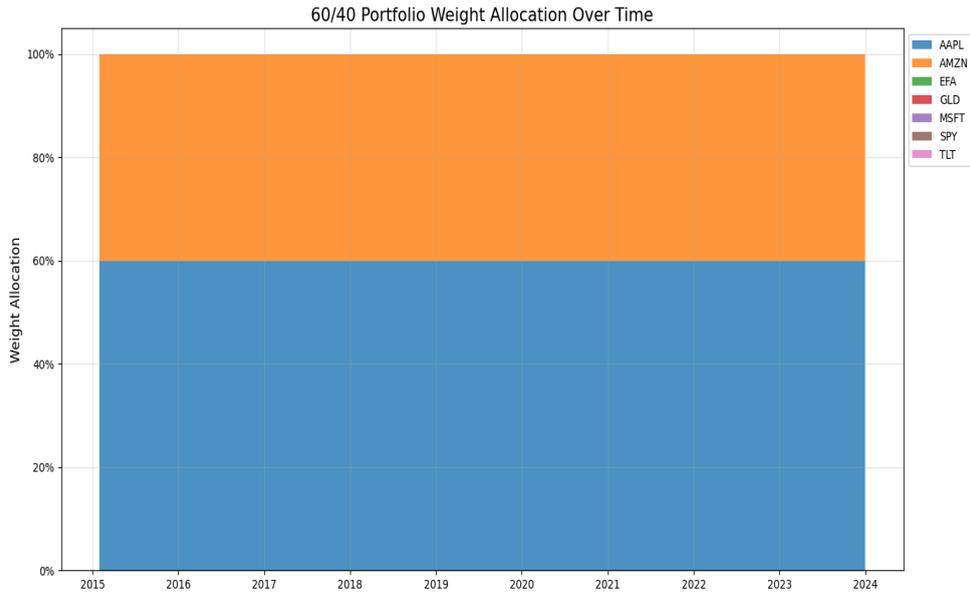

Figure 11. 60/40 Portfolio Weight Allocation Over Time

| Strategy | Cumulative return (%) | Volatility (%) | Sharpe Ratio | CAGR (%) | Max Drawdown (%) |
|---|---|---|---|---|---|
| DRL (PPO) | 50.06 | 15.71 | 0.3671 | 4.62 | -37.36 |
| Autoencoder | 287.79 | 15.45 | 1.0546 | 16.29 | -27.70 |
| Autoencoder+DRL | 273.76 | 15.26 | 1.0390 | 15.81 | -26.52 |
| Transformer+GNN | 269.71 | 14.67 | 1.0659 | 15.67 | -18.81 |
| MVO | 461.31 | 19.38 | 1.0885 | 21.18 | -31.82 |
| Equal-Weighted | 287.79 | 15.45 | 1.0546 | 16.29 | -27.70 |
| 60/40 Portfolio | 332.89 | 21.09 | 0.8798 | 17.72 | -32.64 |

Table 2: Summary of Portfolio Performance

### 4.2 Performance Analysis
**i. Return metrics (Cumulative and Annualized Return)**

The cumulative return and annualized return (see Table 2) reveal the total wealth accumulation and the average yearly growth rate for each strategy. Mean-Variance Optimization investment approach achieved the maximum cumulative return of 461.31% and a CAGR of 21.18%, and this suggests that the classical optimization method performed exceptionally well. Among the deep learning strategies, the Autoencoder+DRL (Cumulative Return: 273.76%, CAGR: 15.81%) and Transformer+GNN (Cumulative Return: 269.71%, CAGR: 15.67%) demonstrated strong performance and this validate the efficiency of deep learning models in detecting and capturing complex market dynamics. The standalone Autoencoder strategy (Cumulative Return: 287.79%, CAGR: 16.29%) and the Equal-Weighted strategy showed identical results. This is because the autoencoder alone did not influence the allocation policy dynamically. These outcomes emphasize that reduction of dimensionality only is not enough reason for return improvement without reinforcement learning for dynamic allocation. The DRL-based, despite being dynamic, underperformed with a CAGR of 4.62%. This highlights the importance of combining DRL with sophisticated feature extraction or structure-aware architectures like Transformers and GNNs.

**Cumulative Return Visualization**

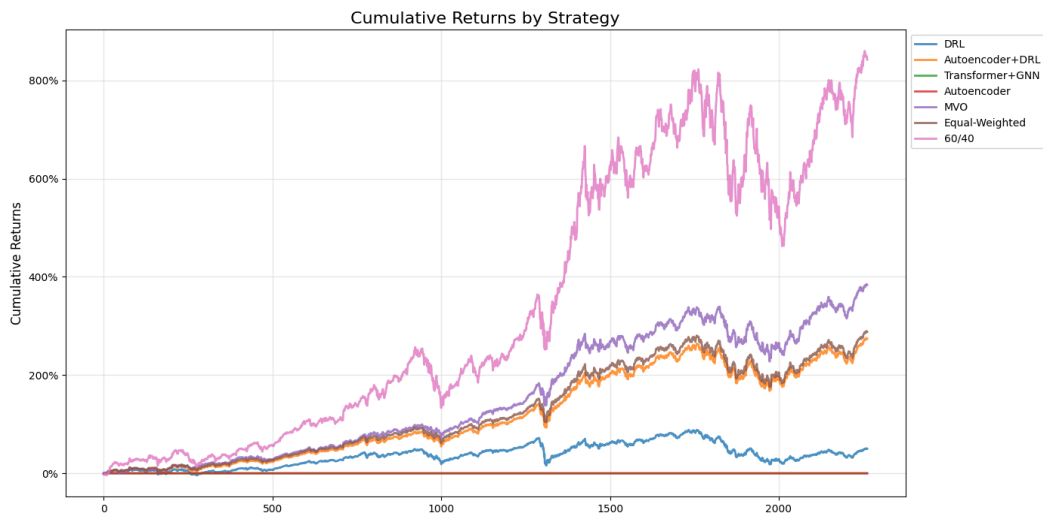

Figure. 11: Cumulative Returns by Strategy

Figure. 11, shows how each total return strategy has grown from the same starting point over a period of time. The 60/40 strategy performed best and showed consistent growth, reaching above 800% cumulative return. MVO, though displays some volatilities, but shows strategic optimization effectiveness. Is the next strong performer having >350% return. Autoencoder displays flat performance and near zero gain. This is because it's only doing feature compression without active decision-making.

**ii    Risk Metrics (Annualized Volatility and Maximum Drawdown)**

Annualized volatility (see Table 2) quantifies the total risk or fluctuation in returns experienced by each portfolio. The volatility levels observed suggest meaningful differences in the risk profiles of the strategies. The Transformer+GNN strategy recorded the lowest volatility (14.67%) and also the lowest maximum drawdown (-18.81%). This suggests superior stability and risk control. It supports the idea that modeling inter-asset dependencies via GNNs can enhance risk-awareness and portfolio robustness. The Autoencoder+DRL and Autoencoder strategies performed well in this regard, with volatility around 15.25% and drawdowns of approximately -26%. The MVO strategy (see Figure. 12), though top in returns, experienced a higher volatility (19.38%) and drawdown (-31.82%). This confirms that higher returns came with higher risk. Static strategies such as Equal Weight and 60/40 showed moderate to high volatility (15.45-21.08%). The strategies also suffered substantial drawdowns (~ - 27% to -32%). These underscore their vulnerability during market downturns due to the absence of adaptive behaviour.

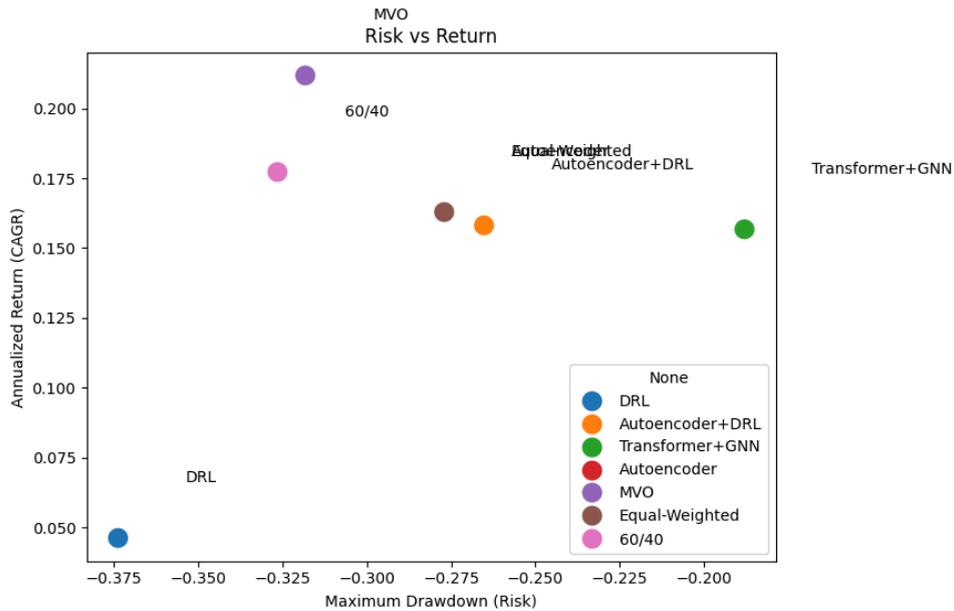

Figure. 12: MVO Risk vs Return

### iii. Risk-Adjusted Return (Sharpe Ratio)

The Sharpe Ratio normalised returns by risk (see Figure. 13) present a different picture. MVO (Sharpe: 1.09) slightly outperformed all other strategies on a risk-adjusted basis. Deep learning strategies follow closely: Transformer+GNN (1.07), Autoencoder (1.05), and Autoencoder+DRL (1.04). This reflects the efficiency of these learning-based methods in generating returns for the amount of risk taken. The DRL baseline lagged significantly with a Sharpe Ratio of 0.37. This shows that deep learning alone does not suffice. A proper architecture and feature extraction are essential.

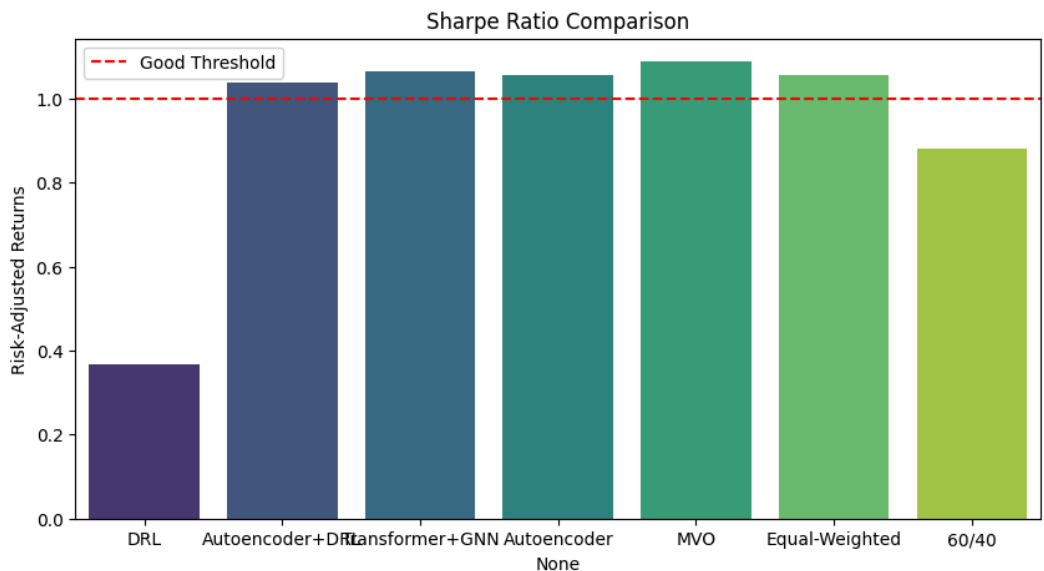

Figure 13: Sharpe Ratio Comparison

### 4.3 Comparison with Existing Literature

These results agreed with findings in the growing literature on data-driven portfolio optimization. Jiang et al. (2017) and Jin et al. (2020) report DRL-based strategies achieving superior Sharpe Ratios and dynamic responsiveness. They also highlight the need for carefully designed architectures. The poor performance of standalone DRL in our case supports their conclusion that naive DRL lacks robustness. The incorporation of an autoencoder aligns with Zhang et al. (2020), who demonstrated that feature compression before DRL enhances learning speed and stability. Our finding supports this, that Autoencoder+DRL outperformed pure DRL in both returns and Sharpe Ratio. The Transformer+GNN strategy builds on the insights from Feng et al. (2020) that GNNs model inter-asset relations to reduce drawdown and improve the Sharpe Ratio. The low volatility and drawdown in our Transformer+GNN model confirm this advantage. The strong performance of MVO diverges from the critique by DeMiguel et al. (2009), who found it often underperforms simple rules like Equal Weight due to estimation errors. Our results suggest that with well-estimated inputs, MVO can remain competitive.

## 5    Discussion of Findings

The empirical results confirm that while deep learning strategies offer strong performance, they do not universally outperform traditional models such as MVO with well-estimated inputs. The Transformer combined with GNN and Autoencoder integrated with DRL strategies exhibited competitive risk-adjusted returns and reduced drawdowns. These architectures demonstrated superior risk management capabilities and more stable learning dynamic behaviour. The DRL baseline, while inherently adaptive, underperformed in the absence of structure-aware inputs. This with existing literature's consensus on architecture sensitivity. The Autoencoder, when applied alone, functioned primarily as a data compressor mechanism rather than a dynamic allocator model. This is the reason why its portfolio allocation closely resembled those of an Equal-Weight portfolio strategy. The Transformer+GNN framework, by modeling both temporal dependencies and inter-asset relationships, proved valuable for smoothing volatility and mitigation of downside risk. This outcome supports the emerging narrative in quantitative finance that relational and sequential modeling plays a critical role in the development of real-world portfolio optimization. Future research may explore the incorporation of deep learning-based feature extraction techniques, such as Autoencoders or Transformers, with the classical portfolio optimization framework. The objective is to enhance decision-making under uncertainty.

## 6.    Conclusion

This research has thoroughly studied multi-faceted models for portfolio optimization. It combines cutting-edge deep learning models such as Deep Reinforcement Learning (DRL), Autoencoders, Transformers, and Graph Neural Networks (GNNs), plus traditional optimization strategies like Markowitz's Mean-Variance Optimization (MVO). A comprehensive methodology involving feature engineering, dynamic rebalancing, risk constraints, and cost-aware strategies was employed. The framework effectively simulates realistic trading conditions. Empirical results demonstrate that hybrid strategies, in particular Transformer+GNN and Autoencoder+DRL, offer superior risk-adjusted returns compared to standalone models. The classical MVO exhibited the highest cumulative returns, deep learning approaches provided enhanced robustness and lower drawdowns, validating their value in risk-sensitive environments. The underperformance of DRL underscores the critical role of structure-aware architectures and informed feature extraction. These findings align with the existing literature and emphasize that no single method universally dominates; rather, the synergy between classical theory and modern machine learning yields optimal performance. This study lays a foundation for future work by exploring hybrid frameworks that integrate domain knowledge with data-driven intelligence in financial decision-making.

# References

Bao, Xiaoyun, Jun Yue, and Yulei Rao. "A Deep Learning Framework for Financial Time Series Using Stacked Autoencoders and Long-Short Term Memory." PLoS ONE, vol. 12, no. 7, 2017, e0180944.

Black, Fischer. "Capital Market Equilibrium with Restricted Borrowing." The Journal of Business, vol. 45, no. 3, 1972, pp. 444–455.

Black, Fischer, and Robert Litterman. "Global Portfolio Optimization." Financial Analysts Journal, vol. 48, no. 5, 1992, pp. 28–43. [https://doi.org/10.2469/faj.v48.n5.28](https://doi.org/10.2469/faj.v48.n5.28).

DeMiguel, Victor, Lorenzo Garlappi, and Raman Uppal. "Optimal versus Naive Diversification: How Inefficient Is the 1/N Portfolio Strategy?" The Review of Financial Studies, vol. 22, no. 5, 2009, pp. 1915–1953. [https://doi.org/10.1093/rfs/hhm075](https://doi.org/10.1093/rfs/hhm075).

Fama, Eugene F., and Kenneth R. French. "The Cross‐Section of Expected Stock Returns." The Journal of Finance, vol. 47, no. 2, 1992, pp. 427–465. [https://doi.org/10.1111/j.1540-6261.1992.tb04398.x](https://doi.org/10.1111/j.1540-6261.1992.tb04398.x).

Feng, Fuli, et al. "Temporal Relational Ranking for Stock Prediction." IEEE Transactions on Knowledge and Data Engineering, vol. 34, no. 1, 2022, pp. 299–313. [https://doi.org/10.1109/TKDE.2020.3006730](https://doi.org/10.1109/TKDE.2020.3006730).

Gu, Shihao, Bryan Kelly, and Dacheng Xiu. "Empirical Asset Pricing via Machine Learning." The Review of Financial Studies, vol. 33, no. 5, 2020, pp. 2223–2273. [https://doi.org/10.1093/rfs/hhaa009](https://doi.org/10.1093/rfs/hhaa009).

He, Guangliang, and Robert Litterman. The Intuition Behind Black-Litterman Model Portfolios. Goldman Sachs Asset Management, 1999.

Heaton, J. B., N. G. Polson, and J. H. Witte. "Deep Learning in Finance." Annual Review of Financial Economics, vol. 9, 2017, pp. 145–181. [https://doi.org/10.1146/annurev-financial-110716-032821](https://doi.org/10.1146/annurev-financial-110716-032821).

https://learn.wqu.edu/my-courses/courses/deep-learning-for-finance

https://learn.wqu.edu/my-courses/courses/portfolio-management

https://engrxiv.org/preprint/download/4355/version/6035/7701/6354

Jiang, B., Y. Xu, and J. Liang. "Stock Movement Prediction with Temporal Graph Attention Networks." ACM Transactions on Knowledge Discovery from Data, 2021.

Jiang, Zhengyao, Dixing Xu, and Jinjun Liang. "A Deep Reinforcement Learning Framework for the Financial Portfolio Management Problem." arXiv, 2017, arXiv:1706.10059. [https://arxiv.org/abs/1706.10059](https://arxiv.org/abs/1706.10059).

Jin, Yichen, Hossam El-Saawy, and Yuying Xu. "Portfolio Management Using Reinforcement Learning: A Systematic Review." Journal of Financial Data Science, vol. 2, no. 4, 2020, pp. 10–30. [https://doi.org/10.3905/jfds.2020.1.048](https://doi.org/10.3905/jfds.2020.1.048).

Ledoit, Olivier, and Michael Wolf. "A Well-Conditioned Estimator for Large-Dimensional Covariance Matrices." Journal of Multivariate Analysis, vol. 88, no. 2, 2004, pp. 365–411. [https://doi.org/10.1016/S0047-259X(03)00096-4](https://doi.org/10.1016/S0047-259X%2803%2900096-4).

Li, Xiaoyue, et al. "Empirical Analysis: Stock Movement Prediction Using Incremental Learning Algorithm and RNNs." Neurocomputing, vol. 347, 2019, pp. 294–305. [https://doi.org/10.1016/j.neucom.2019.02.019](https://doi.org/10.1016/j.neucom.2019.02.019).

Lintner, John. "The Valuation of Risk Assets and the Selection of Risky Investments in Stock Portfolios and Capital Budgets." The Review of Economics and Statistics, vol. 47, no. 1, 1965, pp. 13–37.

Mandelbrot, Benoit. "The Variation of Certain Speculative Prices." The Journal of Business, vol. 36, no. 4, 1963, pp. 394–419.

Markowitz, Harry. "Portfolio Selection." The Journal of Finance, vol. 7, no. 1, 1952, pp. 77–91. [https://doi.org/10.2307/2975974](https://doi.org/10.2307/2975974).

Meucci, Attilio. "The Black-Litterman Approach: Original Model and Extensions." SSRN Electronic Journal, 2008. [https://doi.org/10.2139/ssrn.1117574](https://doi.org/10.2139/ssrn.1117574).

Sharpe, William F. "Capital Asset Prices: A Theory of Market Equilibrium under Conditions of Risk." The Journal of Finance, vol. 19, no. 3, 1964, pp. 425–442. [https://doi.org/10.2307/2977928](https://doi.org/10.2307/2977928).

Wang, Yiming, and Xiaowei Zhou. "Stock Market Prediction via GCNs: Modeling and Analysis." Neural Computing and Applications, vol. 33, no. 18, 2021, pp. 11739–11755.

Ye, Hongyuan, Bin Liu, and Guodong Long. "Reinforcement Learning for Portfolio Optimization with Performance-Driven Reward." Applied Soft Computing, vol. 91, 2020, 106198. [https://doi.org/10.1016/j.asoc.2020.106198](https://doi.org/10.1016/j.asoc.2020.106198).

Zhang, Yujun, Sebastian Zohren, and Stephen Roberts. "Deep Reinforcement Learning for Trading." Journal of Financial Data Science, vol. 2, no. 2, 2020, pp. 25–40. [https://doi.org/10.3905/jfds.2020.1.033](https://doi.org/10.3905/jfds.2020.1.033).

Yahoo Finance. [https://finance.yahoo.com/](https://finance.yahoo.com/).